\documentclass[doublecol, figures]{epl2}
\usepackage{amsmath}
\usepackage{hyperref}
\begin{document}
\title{Spin waves cause non-linear friction}
\author{M. P. Magiera \inst{1}\thanks{E-mail: \email{martin.magiera@uni-due.de}} \and L. Brendel \inst{1} \and
  D. E. Wolf \inst{1} \and U. Nowak \inst{2}}
\shortauthor{M. P. Magiera \etal}
\institute{
\inst{1} Faculty of Physics and CeNIDE, University of Duisburg-Essen,
D-47048 Duisburg, Germany, EU\\
\inst{2} Department of Physics, University of Konstanz, D-78457
Konstanz, Germany, EU
}
\date{\today}
\pacs{75.10.Hk}{Classical spin models}
\pacs{68.35.Af}{Atomic scale friction}
\pacs{75.30.Ds}{Spin waves}
\abstract{
  Energy dissipation is studied for a hard magnetic
  tip that scans
  a soft magnetic 
   substrate.  The 
   dynamics of the atomic moments are 
   simulated by solving the Landau-Lifshitz-Gilbert (LLG) equation
    numerically. The local energy currents are analysed for the
    case of a Heisenberg spin chain taken as substrate. This leads to
    an explanation for the velocity dependence of the friction
    force: The non-linear contribution for high velocities can be
    attributed to a spin wave front pushed by the tip along the substrate.
}
\maketitle
\begin{figure*}
  \onefigure[width=\textwidth]{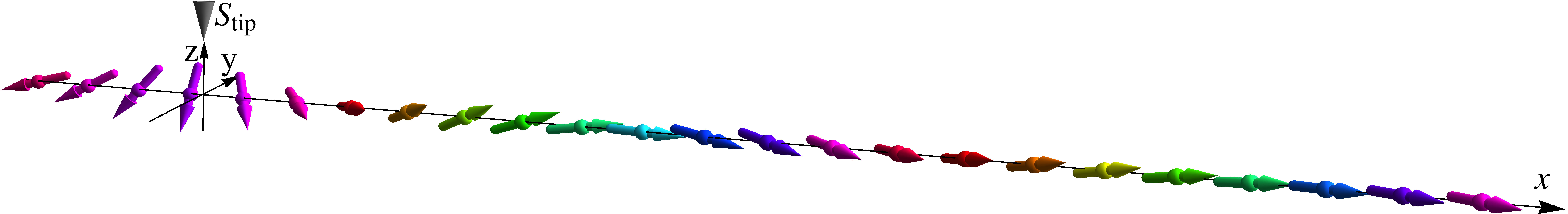}
\caption{
\label{fig:skizze}Snapshot of a simulation. The colour encoding denotes the spins'
  orientation in the $yz$-plane (the tip is moved along the spin
  chain axis to the right). In front of the tip a spin wave is
  visible, \textit{i.e.}\ an oscillation of the spins around the $x$-axis.}
\end{figure*}
\section{Introduction}
On the macroscopic scale the phenomenology of friction is well-known. However,
investigations of energy dissipation on the micron and nanometer scale
have led in recent years to many new insights \cite{Urbakh2004}.
This progress was made
possible by the development of modern surface science methods, in
  particular
Atomic Force Microscopy, which allows to measure energy
dissipation caused by relative motion of a tip with respect to a
substrate.

Studies concerning the contribution of magnetic degrees of freedom to
energy dissipation \cite{Corberi99, Acharyya95} form a young
subfield of nanotribology,
which has been attracting increasing interest in recent years.
Two classes of models have been considered, which show different
  phenomena. The first one is Ising-like spin systems 
  with two equivalent half spaces
  moving relative to each other \cite{Kadau08, Angst10,
    Hucht10, Igloi11, Hilhorst11}. In this case, friction is induced by
  thermal fluctuations, and hence is not present at zero
  temperature. In the second class of models \cite{Fusco08, Paper08,
    Proc09, NIC-Proceeding, Drag_Forces}, there is no symmetry
  between slider and substrate: The slider, representing \textit{e.g.}\ the tip
  of a Magnetic Force Microscope, interacts only locally with a planar
  magnetic surface. While scanning the surface, the tip in general
  excites substrate spins and hence experiences friction, even at zero
  temperature. The present study belongs to the second class of models.

We investigate the nature of the substrate excitations caused by the
tip motion for a classical Heisenberg model with
  Landau-Lifshitz-Gilbert (LLG, \cite{LandauLifshitz1935,Gilbert1955}) dynamics 
(precession around, and relaxation into the local field direction). 
As the spins are continuous variables, spin wave excitations are
possible. 
As we will show in the following,
their properties are reflected in the velocity dependence of
the friction force.
Spin waves are increasingly attracting interest: \textit{e.g.}\ in the last years a new
  subfield of magnetism, \textit{magnonics}, has been developed,
  where materials are studied with respect to their spin wave
  properties \cite{Grundler09, Hille10}. One motivation is to create new devices
  using spin wave logics or novel concepts of data storage.

In a previous work we showed that friction 
in this model 
is proportional to the scanning velocity $v$ (``viscous behaviour''),
provided that the 
tip does not move too fast \cite{Paper08}. The reason can be found in
continuous excitations, while the motion in the Ising-model
consists of discrete excitations and relaxations,
which yields a constant friction force for low $v$. 
In the present paper we focus on the local dissipation
processes in order to explain, why for high velocities deviations from
the viscous behaviour exist. 

\section{Simulation model}
To simulate a solid magnetic material, we consider a chain of
$N{=}320$ classical, normalised dipole moments (``spins'',
\textit{cf.}\ fig.~\ref{fig:skizze})
$\mathbf S_i {=} \boldsymbol \mu_i/\mu_s$, where $\mu_s$ denotes the
material-dependent atomic magnetic moment. 
The spins represent magnetic moments of single
atoms, arranged with a lattice constant $a$ along the $x$-axis. 
Two lattice constants above the spin chain,
a magnetic tip $\mathbf S_\mathrm{tip}$
moves with constant velocity $\mathbf v{=} v {\mathbf e}_x$.
Its magnetisation is fixed in $z{-}$direction.
At the beginning of each simulation
the tip is positioned at the centre of the chain.

In order to keep boundary effects small, we use
a conveyer belt technique with anti-periodic boundary conditions:
When the tip has moved by one lattice constant, the 
boundary spin at the back end is deleted and a new
spin with opposite direction is added at the front end of the chain.
This shift puts the tip back to the centre of the simulation cell.

This paper analyses, how local excitations contribute to magnetic
friction. Each substrate spin contributes its exchange interaction, 
$\mathcal H_\mathrm{sub}^{(i)}$, and its interaction with the tip
field, $\mathcal H_\mathrm{tip}^{(i)}$ to the Hamiltonian
\begin{equation}
\mathcal H = \sum_{i=1}^N \left(\mathcal H_\mathrm{sub}^{(i)} 
+ \mathcal H_\mathrm{tip}^{(i)}\right) = \sum_{i=1}^N \mathcal H^{(i)}.
\end{equation}
As the substrate spins represent a ferromagnetic solid, we use the
anisotropic Heisenberg-Hamiltonian,
\begin{equation}
\mathcal H_\mathrm{sub}^{(i)} = -\frac{J}{2} \mathbf S_i \cdot 
\left (\mathbf S_{i+1}+\mathbf S_{i-1} \right )
 - d_z S_{i,z}^2.
\end{equation}
$J{>}0$ describes the ferromagnetic exchange interaction between
$\mathbf S_i$ and its
nearest neighbours $i{\pm}1$. In order to avoid double counting, half
of the pair interaction is attributed to either spin. 
$d_z{=}{-}0.1J$ is the anisotropy constant: Here it defines an easy plane
anisotropy, thus the substrate spins prefer an alignment in the
$xy$-plane.
The moving tip interacts with each substrate spin by the dipolar
interaction,
\begin{equation} 
\mathcal H_\mathrm{tip}^{(i)} =-w \frac{3 (\mathbf S_i
      \cdot \mathbf e_i) (\mathbf S_\mathrm{tip} \cdot \mathbf e_i) -
      \mathbf S_i \cdot \mathbf S_\mathrm{tip}}{R_i^3},
\end{equation}
where $R_i = \left | \mathbf R_i\right |$ is the length of the distance
vector $\mathbf R_i = \mathbf r_i-\mathbf r_\mathrm{tip}$, and
$\mathbf e_i$ its unit vector $\mathbf e_i=\mathbf R_i/R_i$. $\mathbf
r_i$ and $\mathbf r_\mathrm{tip}$ denote the position vectors of the
substrate spins and the tip, respectively. $w$ quantifies the
dipole-dipole coupling of the substrate and the tip, with 
$w\left| \mathbf S_\mathrm{tip}\right|=10 J a^3$ in this paper.

While the tip magnetisation direction is fixed in time, the substrate spins are allowed to
change their orientation. To simulate their dynamics, we solve
the LLG equation,
\begin{align}
\dot{\mathbf S_i} =&
-\tilde\gamma \left [ \mathbf S_i \times \mathbf
  h_i + \alpha \mathbf S_i \times (\mathbf S_i\times \mathbf h_i)\right ],
\label{eq:LLG}
\end{align}
numerically via the Heun integration scheme,
where $\tilde\gamma = \gamma\ [\mu_s(1+\alpha^2)]^{-1}$ with the
gyromagnetic ratio $\gamma$.
The first term represents the Lamor precession of each spin in the
effective field,
\begin{equation}
\mathbf h_i = -\frac{\partial \mathcal H}{\partial \mathbf S_i}
\end{equation}
with the precession frequency $\tilde\gamma | \mathbf h_i|$. The
precessional motion preserves energy.   
Dissipation is introduced by the second term which causes an
alignment towards $\mathbf h_i$. $\alpha$ is a material constant which can be
obtained from ferromagnetic resonance experiments and represents the
coupling of each spin to a reservoir of zero temperature.
By adding a stochastic term to the effective field, it is possible to study the influence of finite temperatures as done in \cite{Paper08, Proc09}. 
However, in order to analyse the non-equilibrium excitations it is advantageous to suppress thermal spin waves by setting temperature equal to zero in this work.

In order to discuss frictional losses occurring in the system
the {\it global} energy balance was analysed in \cite{Paper08}:
\begin{align}
\frac{d\mathcal H}{dt} =& \  P_\mathrm{pump} - P_\mathrm{diss},
\label{eq:dHdt_global}\\
P_\mathrm{pump} =& \sum_{i=1}^N P_\mathrm{pump}^{(i)} = \sum_{i=1}^N  \frac{\partial \mathcal H_\mathrm{tip}^{(i)}}
{\partial {\mathbf r}_\mathrm{tip}}\cdot \dot{\mathbf
  r}_\mathrm{tip},\\
P_\mathrm{diss} =& \sum_{i=1}^N P_\mathrm{diss}^{(i)}
 = \sum_{i=1}^N \tilde\gamma \alpha
\left (\mathbf S_i \times \mathbf h_i\right )^2. \label{eq:pdiss}
\end{align}
The only explicit time-dependence of the Hamiltonian $\mathcal H$
stems from the motion of the tip. It leads to the first term in
eq.~(\ref{eq:dHdt_global}), which is the energy pumped
into the spin system per unit time by an outside energy source that
keeps the tip moving. Accordingly we call it the
``pumping power''. Its local contribution, $P_\mathrm{pump}^{(i)}$,
is the energy transferred per unit time from the tip to substrate spin
${\mathbf S}_i$.
The friction force, $\mathbf F = - F
  {\mathbf e}_x$, the substrate exerts on the tip is given by
\begin{equation}
F=\frac{\langle P_\mathrm{pump}\rangle}{v},
\label{eq:F}
\end{equation}
 where the angular brackets denote a time average over at
 least one period $a/v$. 

$P_\mathrm{diss}^{(i)}$ represents the energy current from
spin ${\mathbf S}_i$ into the heat bath. In other words, this is the energy
 dissipated at site $i$ per unit time. Dissipation always occurs when
 the system relaxes towards the 
ground-state, in which the spin at
site $i$ is aligned with the local field-direction $\mathbf h_i$.
Without tip movement, $P_\mathrm{pump}$ is zero and
$P_\mathrm{diss}$ leads the system quickly to its
ground state. For a tip moving at constant velocity, a steady
non-equilibrium state is reached, where the time
averaged derivative $\langle d\mathcal H/dt\rangle$ 
vanishes, because all energy pumped into the system is
dissipated. Then the two power terms in eq.~(\ref{eq:dHdt_global})
cancel. 

When evaluating the {\it local} energy balance instead of
  eq.~(\ref{eq:dHdt_global}), energy currents $j_\mathrm{E}$ within the
  substrate have to be taken into account, which transport energy from one spin to its neighbour (cf.\ fig.~\ref{fig:energy_balance}). By taking the time
  derivative of the local Hamiltonian one obtains:
\begin{figure}
\onefigure[width=.49\textwidth]{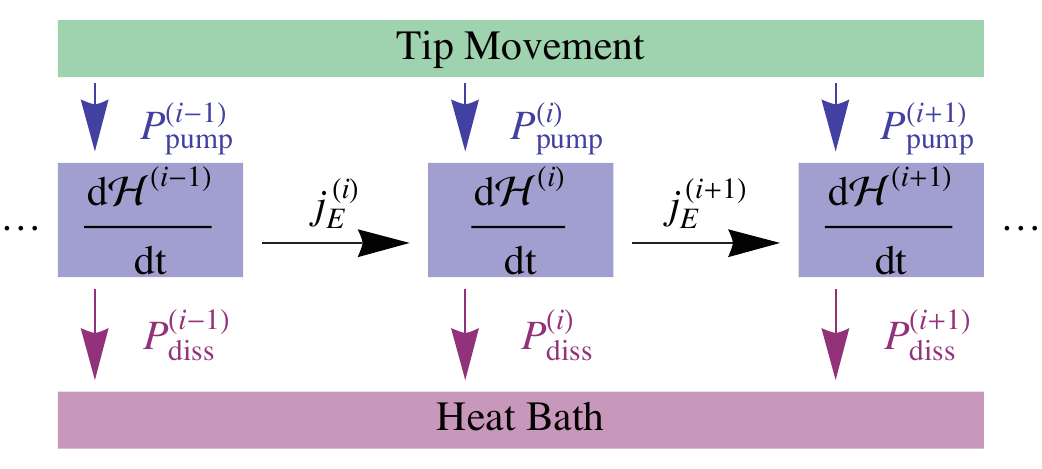}
\caption{\label{fig:energy_balance}Illustration of the local energy balance. The arrows
    represent the directions in which the energy transfers are counted
    positive: a positive $j_E^{(i)}$
   rises the energy at site $(i)$, but lowers the one at site $(i{-}1)$.}
\end{figure}
\begin{align}
  \frac{d\mathcal H^{(i)}}{dt} =& P_\mathrm{pump}^{(i)} - 
  a\;(\mathrm{div}\,j_\mathrm{E})^{(i)} - P_\mathrm{diss}^{(i)} ,
  \label{eq:dHdt}\\
  \notag
  j_\mathrm{E}^{(i)} =& 
  -J({\mathbf S}_{i}-{\mathbf S}_{i-1})\cdot
  \frac{\dot{\mathbf S}_{i-1}+\dot{\mathbf S}_{i}}{2}\\
  =&-\frac{J}{2}(\mathbf S_i\cdot\dot{\mathbf S}_{i-1}-\mathbf S_{i-1}\cdot\dot{\mathbf S}_i)\ .
  \label{eq:jE}
\end{align}

\section{Simulation Results}
\begin{figure}[t]
\onefigure[width=0.4\textwidth]{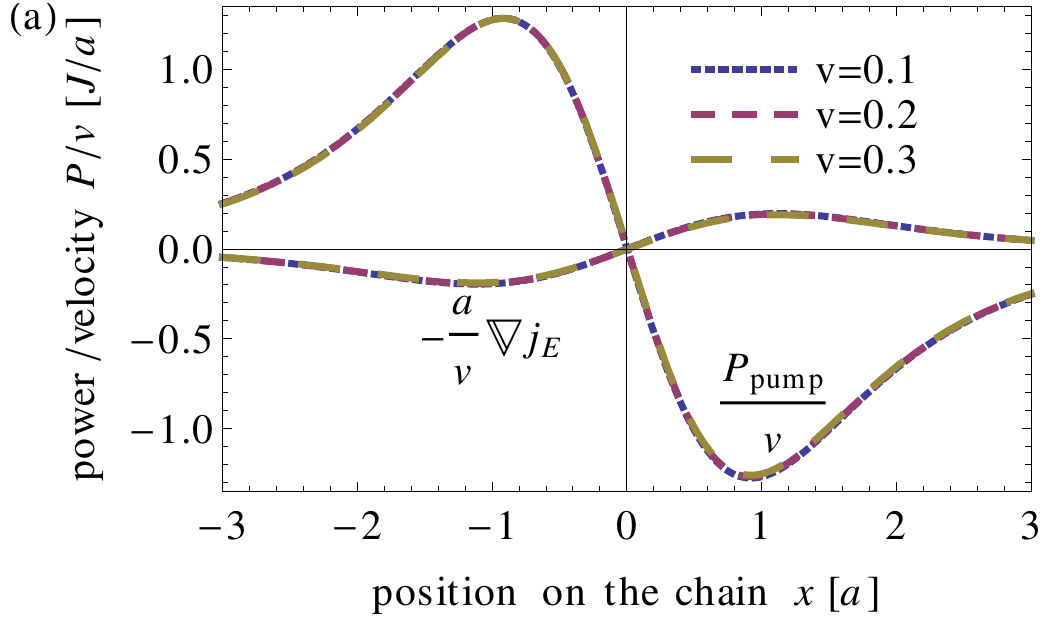}
\onefigure[width=0.4\textwidth]{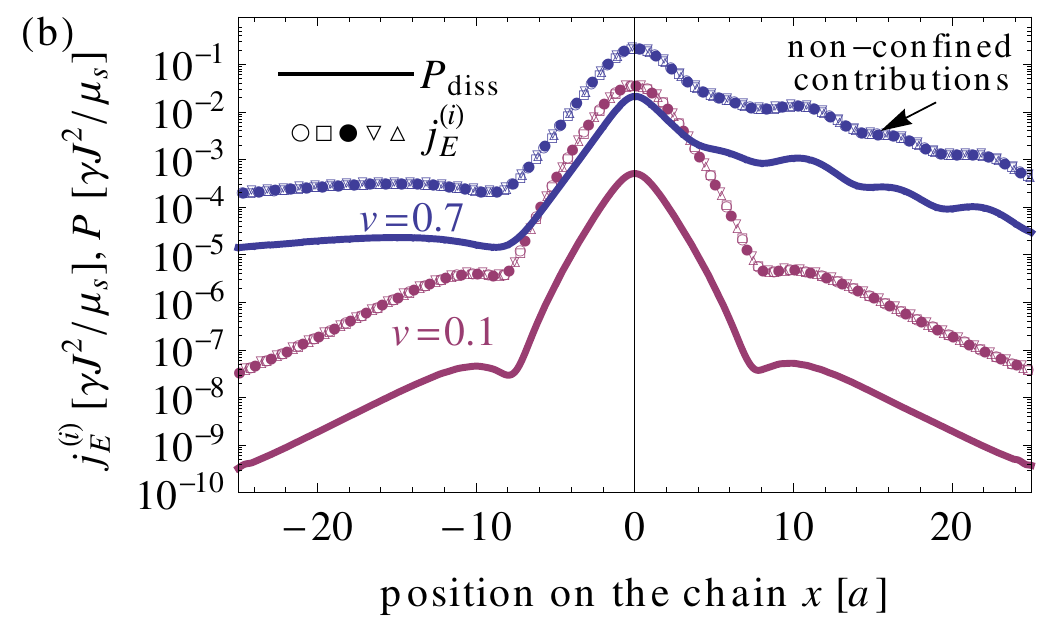}
\caption{\label{fig:powers}(a) Pumping power and transported power,
  rescaled by velocity, at $\alpha=0.1$. Negative pumping power
  represents an energy return from the spin chain to the tip,
  positive an energy injection from the tip into the chain.
(b) Energy current for five different time steps (points, for $\alpha=0.1$),
as well es the dissipation power (solid lines). 
They turn out to be exactly proportional to each other, the
coefficient being linear in $\alpha v$.}
\end{figure}

Let us consider the steady state in a co-moving frame:
the local quantities do not depend on
spin index $i$ and time $t$ separately, but only on the
(continuous) coordinate  $x_i={\mathbf R}_i(t)\cdot {\mathbf
  e}_x$. For considerations, where all spins are equivalent, we can
drop the index $i$, \textit{e.g.}\ the tip position is always at $x =0$. 
In its vicinity, fig.~\ref{fig:powers}(a) shows the local pumping
power, as well as the discrete divergence of the energy current,
$(\mathrm{div}\,j_\mathrm{E})^{(i)}=\big
(j_\mathrm{E}^{(i+1)}{-}j_\mathrm{E}^{(i)}\big )/a$, as functions of
$x$.

The physical interpretation of fig.~\ref{fig:powers}(a) is the following: When the tip
approaches, a substrate spin lowers its energy by adjusting to the
inhomogeneous tip field at the cost of the exchange interaction.  When
the tip has passed by, it returns asymptotically to its higher energy
in the absence of the tip field. This means that the tip injects
energy $P_\mathrm{pump}(x) \propto v$ per unit time at $x<0$ and
extracts apparently the same amount of energy from the
substrate spins at $x>0$. With respect to origin and curve
  shape, this is very similar to an electrical charge
  passing by a charge of opposite sign on a straight line. The
  apparent central symmetry holds only up to first
  order in $v$, though. The small asymmetry, not noticeable in
  fig.\ref{fig:powers}(a), is due to dissipation, which will be
  discussed below. But first we derive the steady state current within
  the chain.

In the steady state, we have
\begin{equation}
  \label{eq:dotS}
  \dot{\mathbf S}_i = \dot{\mathbf S}(x_i) = v\,\partial_x\mathbf S(x_i)
  = v\;\frac{\mathbf S_{i+1}-\mathbf S_i}{a},
\end{equation}
where the third equality, due to using the difference quotient, holds up
to first order in $a$. Plugging that into eq.\ (\ref{eq:jE}), the
current $j_\mathrm{E}^{(i)}$ reads
\begin{equation}
  j_E^{(i)}=\frac{Jv}{2a}(1 - \mathbf S_{i-1} \cdot \mathbf S_{i+1})~.
\label{eq:strom_naeherung}
\end{equation}
This shows that $j_\mathrm{E}$ transports the exchange energy to be
paid for orienting the spins according to the inhomogeneous tip
field. Correspondingly, the source of this current ($\mathrm{div}
j_\mathrm{E}>0$) is behind, and its sink ($\mathrm{div}
j_\mathrm{E}<0$) is in front of the tip, as seen in
fig.~\ref{fig:powers}(a).

The terms discussed so far are reversible and hence independent of
  the damping constant $\alpha$: To first order in $v$ they add up to zero in
  eq.~(\ref{eq:dHdt_global}). The origin of dissipation is that the
  spin pattern does not follow the tip instantaneously, but with a
  delay, which corresponds in the co-moving frame to a lag $\Delta x \propto \alpha v$ \cite{Paper08}. It is a manifestation of the
  \textit{driving} out of equilibrium, which the spin
  relaxation must counteract. Fig.~\ref{fig:powers}(b) shows
  that $P_\mathrm{diss}(x)/j_\mathrm{E}(x)$ is indeed a
  constant $\propto\alpha v$, which we may call the \textit{driving force}\footnote{This is analogous
    to electrical power $P = U I$, where the voltage $U$ provides the
    driving force for the current $I$.}. As pointed out in 
  fig.~\ref{fig:powers}(a), $j_\mathrm{E}$ is proportional to
  $v$. This only holds true for velocities not much larger than $v_0\approx
  0.31 \gamma J a/\mu_s$ (see eq.~(\ref{eq:wavenumber}) below). This
  implies that for 
  small velocities $P_\mathrm{diss}\propto\alpha v^2$, which gives rise to a
  friction force, eq.~(\ref{eq:F}), proportional to $\alpha v$.

\begin{figure}[tb]
\onefigure[width=.4\textwidth]{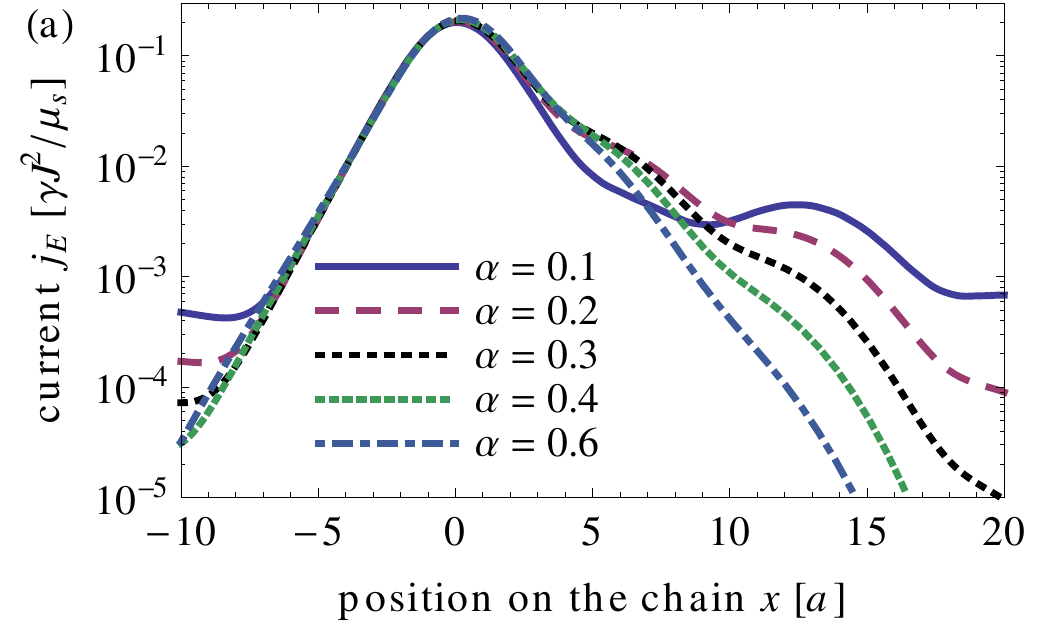}
\onefigure[width=.4\textwidth]{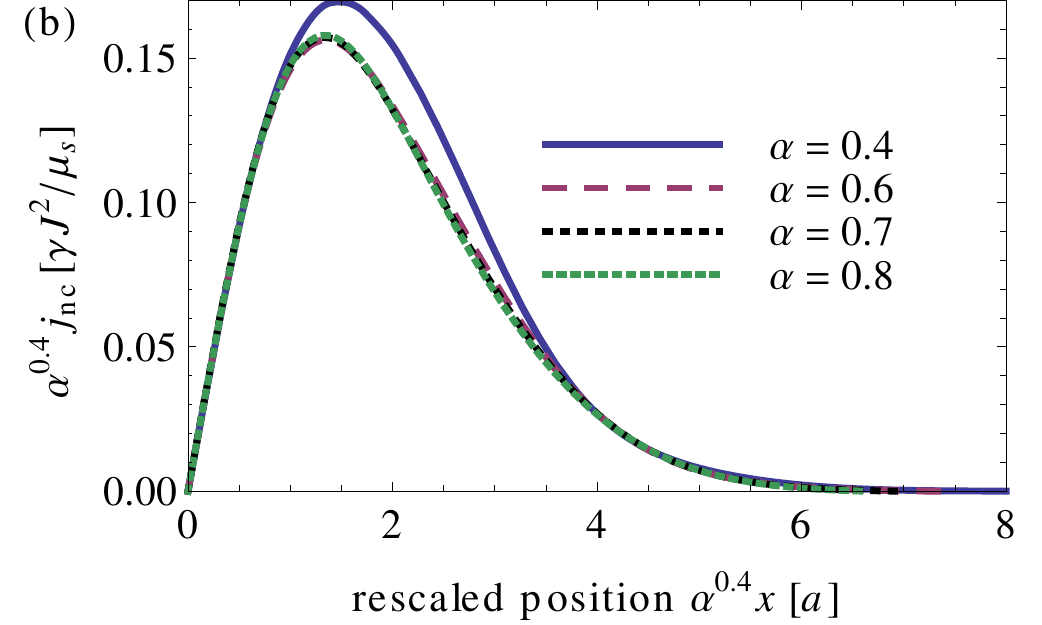}
\caption{\label{fig:range}(a) Energy currents for $v{=}0.6$ and some $\alpha$ values. When the
damping is lowered, the current can proceed further in the substrate.
(b) Rescaled non-confined current vs.\ a rescaled position for
$v{=}0.85$.
}
\end{figure}

Fig.~\ref{fig:powers}(b) shows an important qualitative difference
between energy currents for tip velocities below, respectively
above $v_0$. For low velocities, the energy current is concentrated
around the tip position in an essentially symmetric way: Whatever
exchange energy is released behind the tip, is reabsorbed in front of
it.
For high velocities, however, an additional shoulder in front
of the tip appears. This shoulder represents a part of the energy current,
which can leave the tip's immediate neighbourhood
and propagates further along
the spin chain, until it is damped out. We call this contribution 
\textit{non-confined}. The propagation
range depends on the damping constant $\alpha$, as can be
seen in fig.~\ref{fig:range}. The lower the damping constant, the
farther the current extends.

In order to evaluate this quantitatively, we define
the non-confined energy current as
\begin{equation}
j_\mathrm{nc}(x) = j_\mathrm{E}(x) - j_\mathrm{E}(-x) \;\;\;\mathrm{for}\;\;\; x{>}0.
\end{equation}
It is plotted for several $\alpha$-values in
fig.~\ref{fig:range}(b). The axes are rescaled in order to show that
the range shrinks with increasing damping approximately like
$\alpha^{-0.4}$, and that the amplitude of the non-confined
current also decreases roughly like $\alpha^{-0.4}$. The range  and
the amplitude of the non-confined current combine in such a way, that
the integral over $j_\mathrm{nc}(x)$ is nearly
proportional to $\alpha^{-1}$. Hence, when multiplied by the driving
  force $\propto \alpha v$, the $\alpha$-dependence nearly
  cancels. The contribution of the non-confined excitations to
friction is therefore approximately independent of $\alpha$, in
  contrast to the confined contribution discussed above. As
  will be explained below, the two contributions also depend
  differently on velocity.

\begin{figure}[t]
\onefigure[width=.48\textwidth]{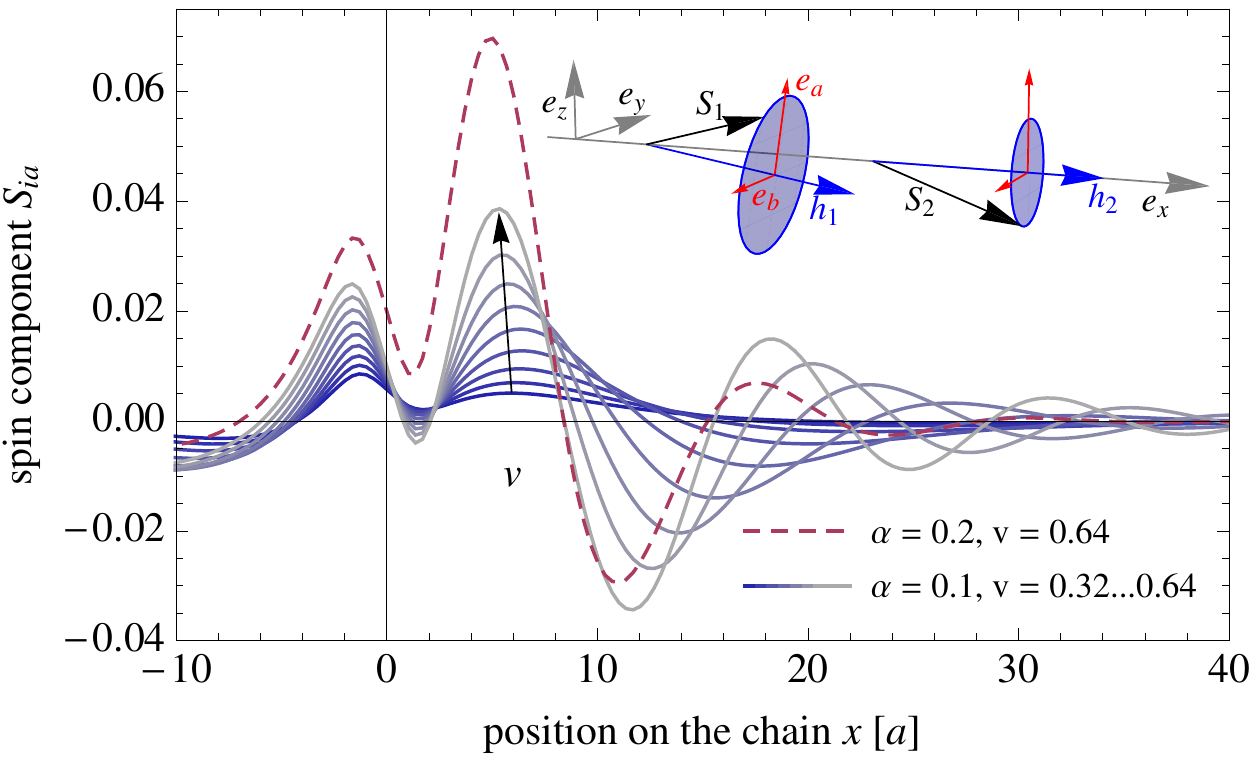}
\caption{\label{fig:spinwaves}Spin component perpendicular to the corresponding
  field. While for low velocities one precession around the local field is
  observable, for higher velocities more precession cycles in front of the tip
  are present. For lower damping excitations
  reach further.
Sketch: Definition of the field dependent basis.
For two sites spin and field values are sketched. For $S_2$ the local field
points in $x-$direction (as the situation in the studied system in
front of the tip is), and the spin precesses on the blue disk. The
shift of the field for $S_1$ to the bottom (as it may be induced by
the tip-field) yields also a change of the disk and the
appropriate basis.
}
\end{figure}

The non-confined excitations can be regarded as spin waves.
An excitation means a deflection of a spin $\mathbf S_i$ from the local
field $\mathbf h_i$, leading to dissipation at the corresponding site
according to eq.~(\ref{eq:pdiss}) and a precession around $\mathbf
h_i$ in a plane perpendicular to $\mathbf h_i$. 
Accordingly, 
\begin{equation}
  \mathbf  e_{i, a}=
  \frac{\mathbf h_i {\times}{\mathbf e}_y}{
  |\mathbf h_i {\times}{\mathbf e}_y|}
  \quad\text{and}\quad
  \mathbf e_{i, b}=
  \frac{\mathbf h_i {\times} \mathbf e_{i, a}}{
  |\mathbf h_i {\times} \mathbf e_{i, a}|}
\label{eq:locbase}
\end{equation}
form an appropriate local basis to illustrate the excitations. 
Spins far in front of the tip
experience a field which points in $x$-direction, thus 
the $\mathbf e_{i, a}$-component points in $z$-direction. 
Near the tip the
basis changes as sketched 
in the inset of fig.~\ref{fig:spinwaves}. 

In the low velocity regime, a deflection from the
local fields is present solely in the vicinity of the tip, according
to the purely confined contribution to friction discussed above. 
For velocities, where non-confined currents can be observed,
additional oscillations are present in front of the tip.
A Fourier analysis of our simulation data shows that their wavenumber has approximately a linear velocity dependence:
\begin{equation}
k \propto v{-}v_0\;\;\; \mathrm{with}\;\;\; v_0\approx0.31~\gamma J
a/\mu_s.
\label{eq:wavenumber}
\end{equation}
The resulting empirical coefficient 0.31 must be expected to depend on system parameters like the tip field's shape and amplitude, which determine the spin waves' confinement.
For small enough damping, the oscillations presumably extend arbitrarily
  far in front of the tip. Asymptotically we can neglect the tip field
  and consider an isolated spin chain with exchange
  interaction only. 
In its ground state, all spins point into the same direction, say
$\mathbf{e}_x$. For small perturbations of $\mathbf S_i$ from this direction,
the LLG-equation with
\begin{equation}
  \mathbf h_i{=}J\,(\mathbf S_{i-1}{+}\mathbf S_{i+1})
  \label{eq:Jonly}
\end{equation}
can be linearised (\textit{e.g.}~\cite{Coey10}). The solution is a
spin wave with an oscillating part of
\begin{equation}
  \label{eq:wave}
  \boldsymbol\delta_i = \mathbf e_y\,\delta\,\cos(ika-\omega t)+\mathbf e_z\,\delta\,\sin(ika-\omega t)
\end{equation}
to first order in its small amplitude $\delta$. Its dispersion relation
\begin{equation}
  \omega(k)=4\frac{J\gamma}{\mu_s}(1{-}\cos(k a))
  \label{eq:dispersion}
\end{equation}
yields a group velocity $v\propto k$ in the long wavelength
limit. The finding eq.~(\ref{eq:wavenumber}) indicates that this holds
true even in the more complicated system with the inhomogeneous tip
field.

Inserting $\mathbf S_i=\mathbf e_x+\boldsymbol\delta_i$ into
eqs.~(\ref{eq:Jonly}) and (\ref{eq:pdiss}) yields for the wave of wave
number $k$ a dissipation of
\begin{equation}
P_\mathrm{diss}(k) \propto \sin^4{\left (\frac{ka}{2} \right )}. 
\end{equation}
Using eq.~(\ref{eq:wavenumber}) and assuming that the amplitude of the
spin wave excitations $\delta$ does not depend on $v$ and that $ka{\ll}1$, 
this predicts a non-linear velocity dependence of the spin wave contribution
to friction like $\frac{(v-v_0)^4}{v}$. 

\begin{figure}[tbh]
\onefigure[width=.4\textwidth]{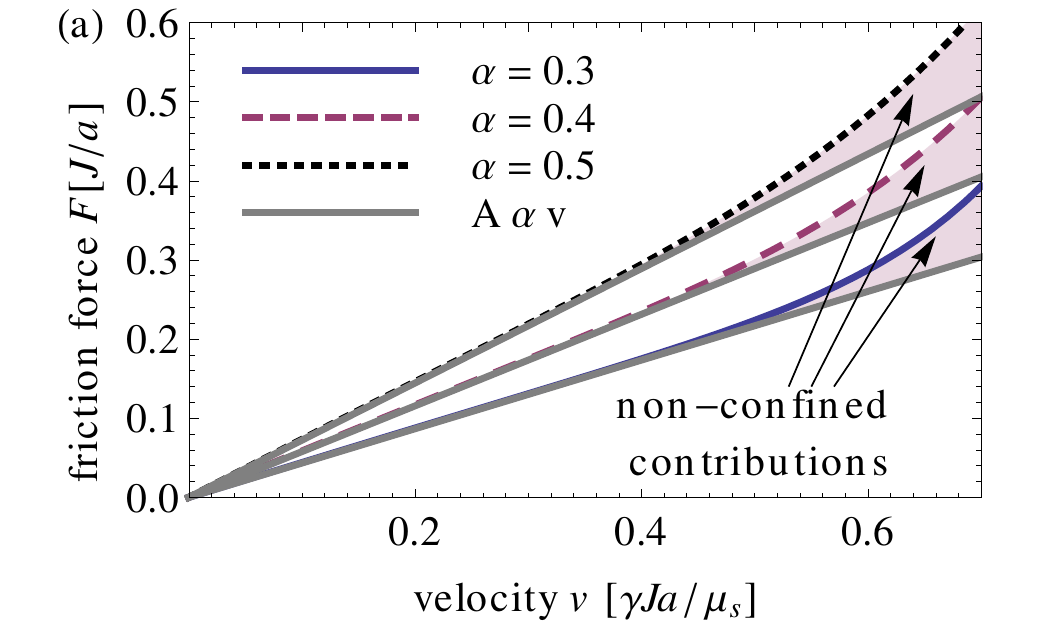}
\onefigure[width=.4\textwidth]{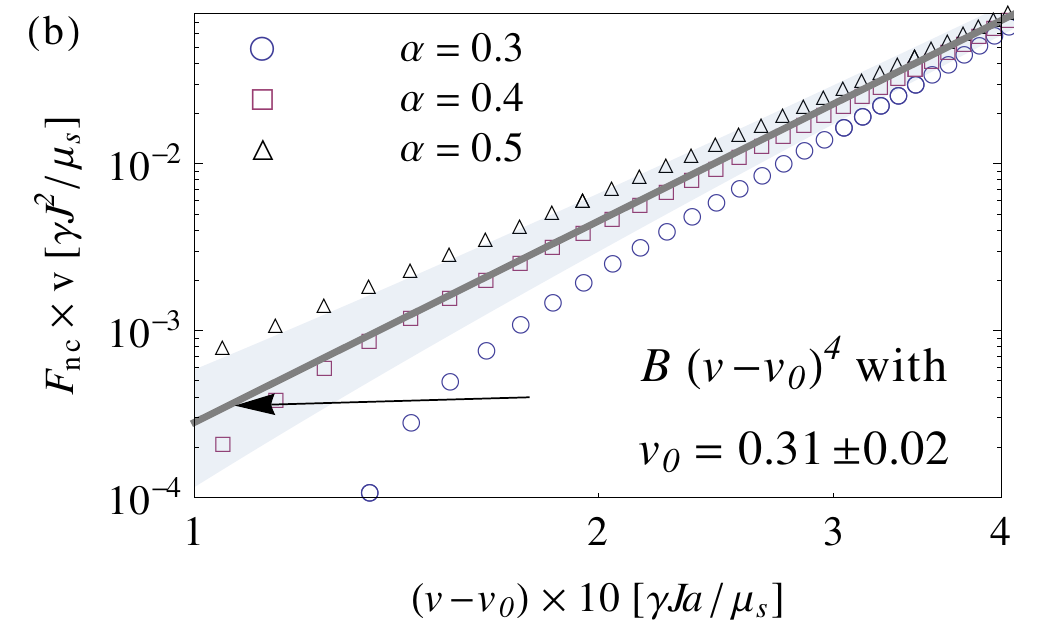}
\caption{\label{fig:friction}(a) Friction force for several velocities,
  with the non-confined part marked by the shading.
(b) Non-confined contributions to the friction force times velocity as
well as the function $B(v-v_0)^4$, where $v_0$ has been taken from eq.~(\ref{eq:wavenumber}).}
\end{figure}

The total magnetic friction force is thus predicted to be
\begin{equation}
F \approx A \alpha v+B(\alpha) \Theta(v-v_0) \frac{(v-v_0)^4}{v},
\label{eq:fric}
\end{equation}
where $\Theta(x)$ is the Heaviside step
function, and the coefficients $A$, $B$ as well as $v_0$ may depend on 
system parameters like the tip field.
In fig.~\ref{fig:friction} this total force is plotted, and the
simulation results are in good agreement with
eq.~(\ref{eq:fric}).

\section{Conclusion and outlook}

In this work, we could separate two distinct contributions to magnetic
friction by examining the energy current in a spin chain.
The confined current results in a friction force 
$F{=}A\alpha v$, which is in accord with our earlier results for
$2d$ \cite{Paper08} and $3d$
\cite{NIC-Proceeding} substrates.
Above a threshold velocity $v_0$, spin wave excitations may leave
  the tip's immediate 
neighbourhood and form a damped wave packet in front of the tip,
propagating along with it.
These excitations are the stronger, the weaker the damping.
They lead to an
additional contribution to friction with a non-linear velocity dependence.
The dependence of the non-confined contribution on $\alpha$ is not trivial,
because the range as well as the amplitude of the energy current are
influenced in a non-linear way., cf.\ fig~\ref{fig:range}(b).

Important extensions of the present investigation include the
influence of dimensionality on the non-confined spin waves. 
Here the propagation is not confined to the tip's motion
direction. The influence of thermal spin waves and their interaction
with the free spin waves is another open question.
Studies dealing with this are already in progress and will
be reported in a future work.

\begin{acknowledgments}
This work was supported by the German Research Foundation (DFG) via
SFB 616 and the German Academic Exchange Service (DAAD) through the PROBRAL programme.
\end{acknowledgments}

\bibliographystyle{eplbib}
\bibliography{paper}
\end{document}